# SOLUBILITY LIMITATION OF METHYL RED AND METHYLENE BLUE IN MICROEMULSIONS AND LIQUID CRYSTALS OF WATER, SDS AND PENTHANOL SYSTEMS

**D. Beri[a], A. Pratami[a], P. L. Gobah[a], P. Dwimala[a], A. Amran[a]**

[a]*Laboratory of Material Science, Department of Chemistry, Faculty of Mathematics and Science, Universitas Negeri Padang, Kampus UNP Air Tawar, Jl. Prof. Dr. Hamka, 25131 Padang, Indonesia*

*e-mail:* deski.beri@gmail.com

**Abstract –** Solubility of dyes in amphiphilic association structures of water, SDS and penthanol system (i.e. in the phases of microemulsions and liquid crystals) was attracted much interest due to its wide industrial and technological applications. This research was focused on understanding the solubility limitation of methyl red and methylene blue in microemulsion and liquid crystal phases. Experimental results showed that the highest solubility of methyl red was in LLC, followed by w/o microemulsion and o/w microemulsion, respectively, whereas the highest solubility of methylene blue was in w/o microemulsion, followed by o/w microemulsion and LLC, respectively. Hence, a chemical dynamics strongly played an important role in the solubility limitation of methyl red and methylene blue in microemulsions and liquid crystal phases.

## INTRODUCTION

The wide applications of surfactants in research and technology lead to the advanced study in the field of physical chemistry and material sciences[1]. Some thermodynamical aspects in terms of solubility has been studied and rapidly developed in this few decades[2]. Since it has many industrial and technological applications, such as: ink[3], paint[4], oil recovery[5], pharmaceutical[6], agricultural[7] and electronic industries[8]. Therefore, by understanding this field the chemical attractions and repulsions, or interactions could be open widely and the "grail" of chemical nature could be expressed in the simple meaning[9]. The aim of this research is to express the limitation of dyes solubility in microemulsions and liquid crystals of water, surfactant and cosurfactant system[10]. We focused on the three components system due to the applicability of such systems in many industries and technologies[11].

Solubility of dyes in microemulsions and liquid crystals could be used to probe the microstructure of colloids association structures[12]. The homogeneity mixtures of microemulsions and liquid crystals systems[13] in macroscopic overview would bring a unique ability to dissolve dyes[14]. This phenomenon could be investigated at molecular level using *ab-initio* and chemical dynamics computations[15]. All interaction at molecular level could be expressed genuinely, however, in this research we proposed the solubility limitation of dyes in microemulsions and liquid crystals(especially in LLC) of water, sodium dodecyl suphate and penthanol system.

# EXPERIMENTAL SECTION

## *Materials and Methods*

The following chemicals were used without further purification: nononic surfactant, SDS (sodium dodecyl sulphate) GR, nitric acid (70%) fuming, potassium hydroxide GR, n-penthanol and methylene blue were purchased from Merck KgaA, Germany, methyl red was obtained from Wako Pure Chemical Industries Ltd. The water was double distilled water (Rafi Medika) In addition, nitric acid (37%) was used as acid medium to adjust the water at pH = 4.5, and potassium hydroxide was used as the base medium to adjust the water at pH = 9.5.

## *Phase Diagram Preparation*

The o/w and w/o microemulsion regions, as well as lamellar liquid crystal (LLC) and hexagonal liquid crystal (HLC) regions [from the water(pH of 4.5 and 9.5), SDS, and penthanol system], were determined by titration of SDS, penthanol with water to turbidity at room temperature ($25^o$ C). The solubility region was checked by long time observation, monitoring the turbidity of the samples both inside and outside the single phase regions. The phases in equilibrium with the microemulsion were identified as lamellar liquid crystal (LLC) and hexagonal liquid crystal (HLC) from its pattern when viewed between crossed polarizers in optical microscopes.

## *Solubility of Methyl Red and Methylene Blue in Microemulsions and Liquid Crystals*

Solubility of methyl red, especially, in o/w and w/o microemulsions, lamellar liquid crystal regions, as well, were determined in the system of water (pH=4.5), SDS and penthanol. Whereas, solubility of methylene blue in o/w and w/o microemulsions and lamellar liquid crystals, was determined in the system of water (pH=9.5), SDS and penthanol. In fact, the solubility of methyl red and methylene blue in HLC relatively low due to one dimensional periodicity structure. The procedure were: small quantity of methyl red/methylene blue were added gradually into a tube that was already filled with/by given composition of samples of interest. Stirring it, using vertexer. Hence, the solubility of methyl red/methylene blue was observed visually. Red and blue laser light were used to convince the visual appearance. In addition, the Hund Wezlar® Optical Polarizing Microscope was also used whenever needed. The addition was stopped when a very little precipitated of dyes showed, and saturated mixture was achieved.

# RESULT AND DISCUSSION

## *Phase Diagram*

Phase diagram of water (both of pH=4.5 and pH=9.5); SDS and penthanol system exhibits four main phase areas, i.e., o/w and w/o microemulsions, lamellar liquid crystal (LLC) and hexagonal liquid crystals (HLC), as shown in **Figure.1.**

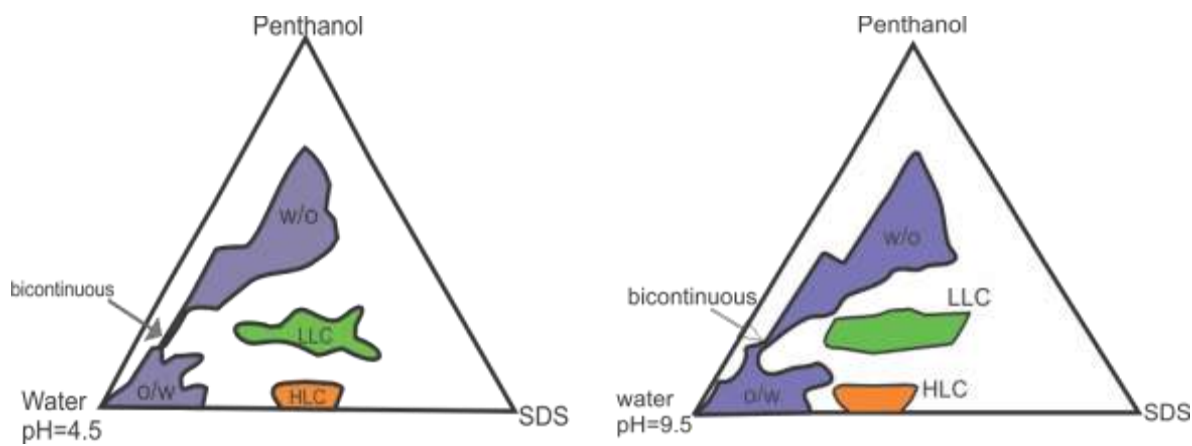

**Figure. 1. Phase diagram of water, SDS and penthanol system which represents four phase areas; o/w and w/o microemulsions, lamellar and hexagonal liquid crystals. The bridge of w/o and o/w microemulsions lays bicontinuous phase. System of water, pH=4.5 (*left*) and system of water pH=9.5 (*right*).**

In water at pH=4.5; SDS and pentanol systems, w/o microemulsion exhibited wide area in the middle of triangle along water and pentanol rich components. The coordinate positions were around 12-60% water; 6-35% SDS; and 30-70% penthanol content. Whereas o/w microemulsion exhibited the finger shape at water rich composition. The coordinate positions were around 70-100% water; 0-25% SDS; and 0-14% penthanol content. In the middle of w/o and o/w microemulsions was a certain structure which united the two structures as a bridge which called bicontinueos phase. This is a unique structure due to the o/w and w/o microemulsions mixed together to form a combination structure. Lamellar liquid crystal (LLC) was formed at the coordinate around 23-55% water; 24-60% SDS; and 14-25% penthanol content. Whereas hexagonal liquid crystal (HLC) was formed at the coordinate around 38-55% water; 40-57% SDS; and 1-7% penthanol content. The complete picture could be seen at the left side of **Figure.1**.

In water at pH=9.5; SDS and pentanol system, w/o and o/w microemulsions exhibited almost similar trends as well as water at pH=4.5. For w/o microemulsion, phase area lays on the middle of the triangle along water and pentahol rich components. The coordinate position were around 12-60% water; 6-50% SDS; and 30-77% penthanol content. Whereas o/w microemulsion exhibited the finger shape at water rich composition. The coordinate position were around 63-100% water; 0-26% SDS; and 0-14% penthanol content. In the middle of w/o and o/w microemulsions lays bicontinous phase as well. Lamellar liquid crystal (LLC) was formed at the coordinate around 25-60% water; 19-49% SDS;

and 18-28% penthanol content. Whereas hexagonal liquid crystal (HLC) was formed at the coordinate around 47-66% water; 28-48% SDS; and 0-8% penthanol content. The complete picture could be seen at the right side of **Figure.1**. Chemical structures of water, SDS and penthanol and cartoonic structures were represented in **Figure.2.**

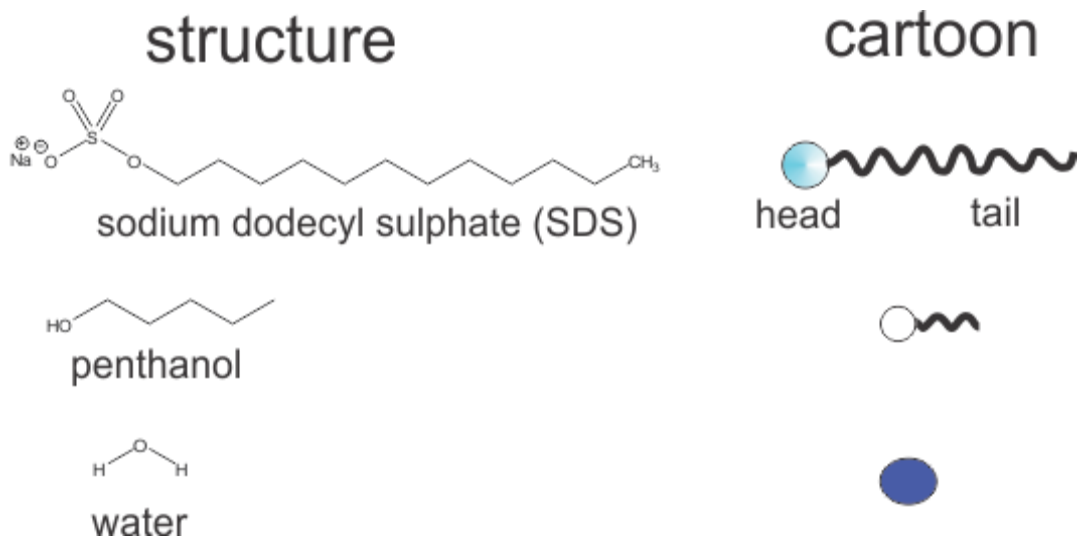

**Figure. 2. Representation of chemical and cartoonic structures of SDS, penthanol and water. All three components arranged in given weight composition to formed phase diagram.**

Both w/o and o/w microemulsions exhibited homogeneous clear solution which were stable in the long time periods at room temperature. The homogeneity was clarified using Hund-Wetzlar® optical polarizing microscope. Under microscope, the solutions was so clear with micro-spherical-dot (**msd**) distributed homogeneously throughout the solutions. The appearing of particles **msd** were measured in 40 μm size in diameter. w/o microemulsion formed at rich oil content (poor water), whereas o/w microemulsion formed at poor oil content (rich water) in the triangle or phase diagram. Under polaroid-parafilm, both w/o and o/w microemulsions were innert in terms of light polarity rotation.

LLC exhibited homogeneous clear high viscous texture/subtance which was stable under long period of time at room temperature. The rheology was measured under Hund-Wetzlar® optical polarizing microscope and it formed a layer structures. LLC formed a series of **msd** horizontally and there was a space among the layers. In addition, HLC exhibited a two dimentional and highly homogeneous viscous texture/substance which was difficult to mixed under thermolyne® mixer. The difficulty was not easy to resolve. Under parafilm apparatus, the HLC phase was rotated the polarization light. The representative formation of o/w and w/o microemulsions could be simplified in cartonic form as shown in **Figure. 3.**

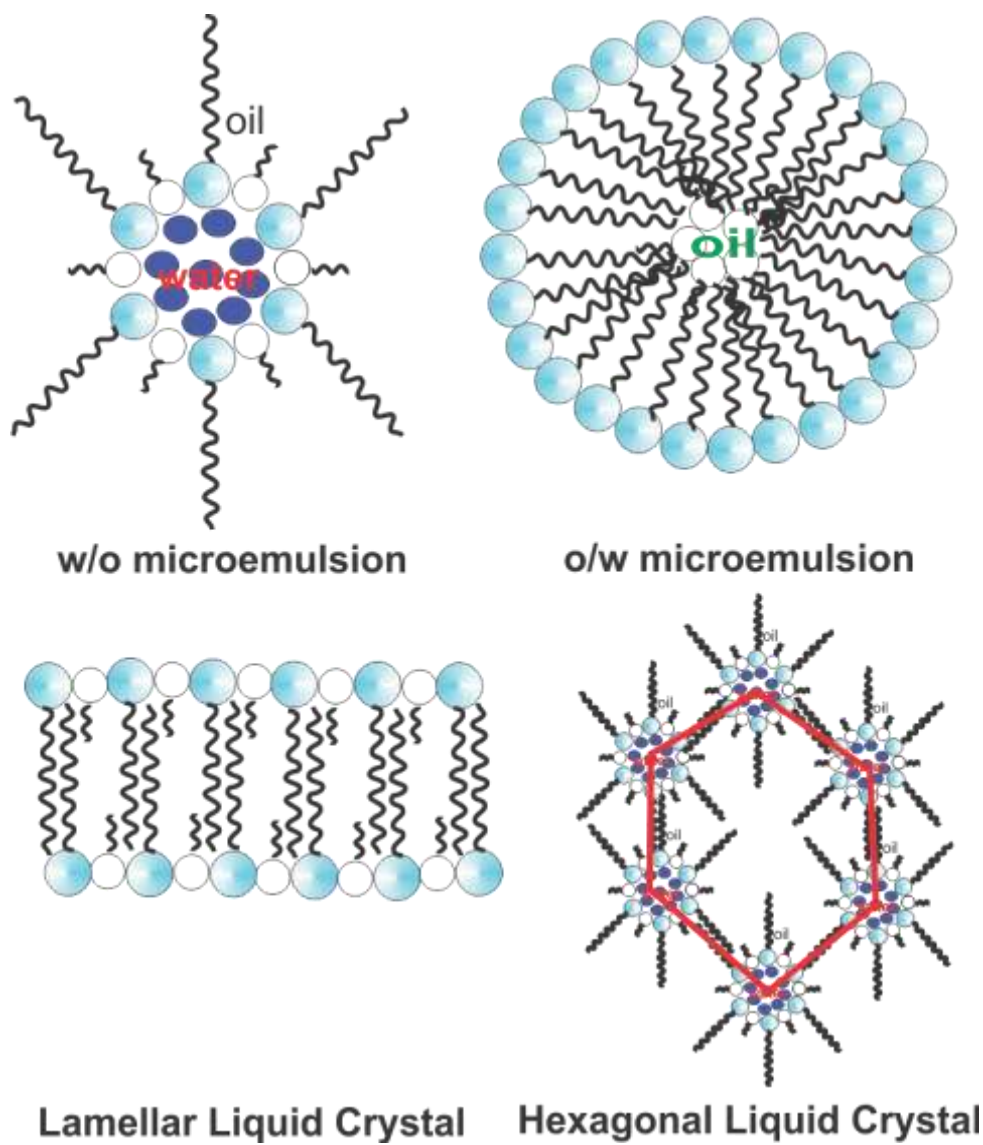

**Figure. 3. Cartoonic representation the two dimensional cross section of w/o and o/w microemulsions and lamellar and hexagonal liquid crystals. Experimental model for phase formation of phase diagram.**

## *Solubility of Methyl Red and Methylene Blue in Microemulsions and Liquid Crystals*

Solubility of methyl red was determined for w/o and o/w microemulsions and LLC phase for system of water at pH=4.5, SDS and penthanol, whereas solubility of methylene blue was determined for w/o and o/w microemulsions and LLC phase for system of water at pH=9.5, SDS and penthanol. We notice that, HLC phase for both system was not determined due to the technical problems. The result was tabulated in **Table 1**.

**Table 1. Solubility Data for Methyl Red in Microemulsions and Lamellar Liquid Crystal of Water pH=4.5, SDS and Penthanol System**

| Phases | Solubility of Methyl Red in mg |
|---|---|
| O/W microemulsion | 0.06 |
| W/O miroemulsion | 0.13 |
| LLC | 0.70 |

The solubilisation of methyl red in LLC was the highest compared to w/o and o/w microemulsions in the water at pH=4.5, SDS and penthanol system (data **Table 1**). In contrast, the solubilisation of methylene blue in LLC was the lowest compared to o/w and w/o microemulsions (data **Table 2**).

**Table 2. Solubility Data for Methylene Blue in Microemulsions and Lamellar Liquid Crystal of Water pH=9.5, SDS and Penthanol System**

| Phases | Solubility of Methyl Red in mg |
|---|---|
| O/W microemulsion | 0.38 |
| W/O miroemulsion | 0.46 |
| LLC | 0.27 |

The chemical structure of methyl red and methylene blue and cartoonic structures represented in Figure. 4.

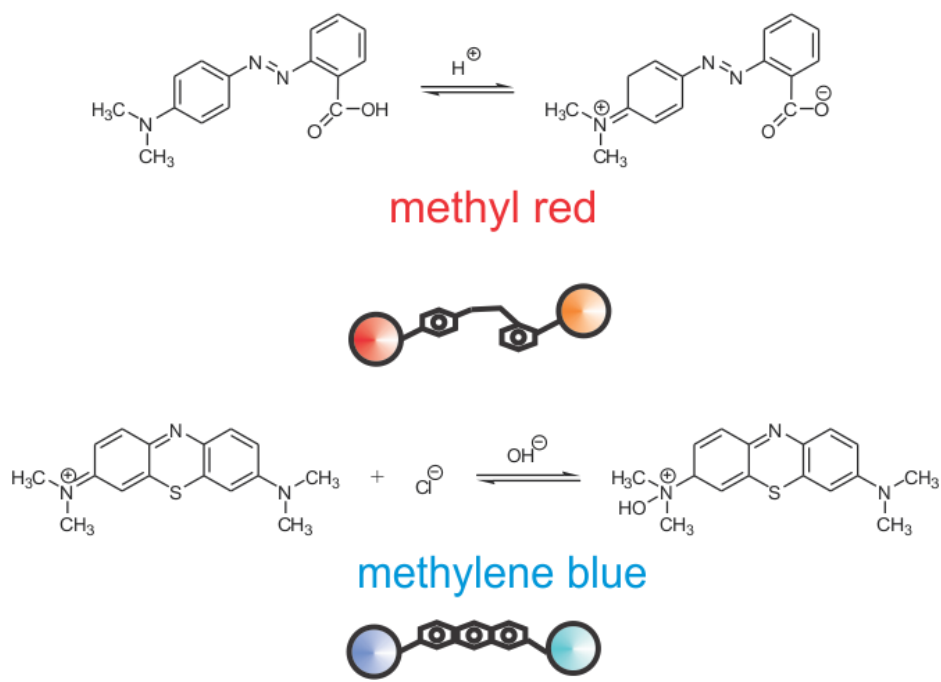

**Figure. 4. Chemical and cartoonic structures for methyl red (*upper*) and methylene blue (*lower*) in their chemical properties condition (acid and basic).**

Using such model as represented in **Figure.4** we could made a model for chemical interaction for methyl red and methylene blue in the association structures. This interaction model could be pictured as **Figure.5** and **Figure.6**.

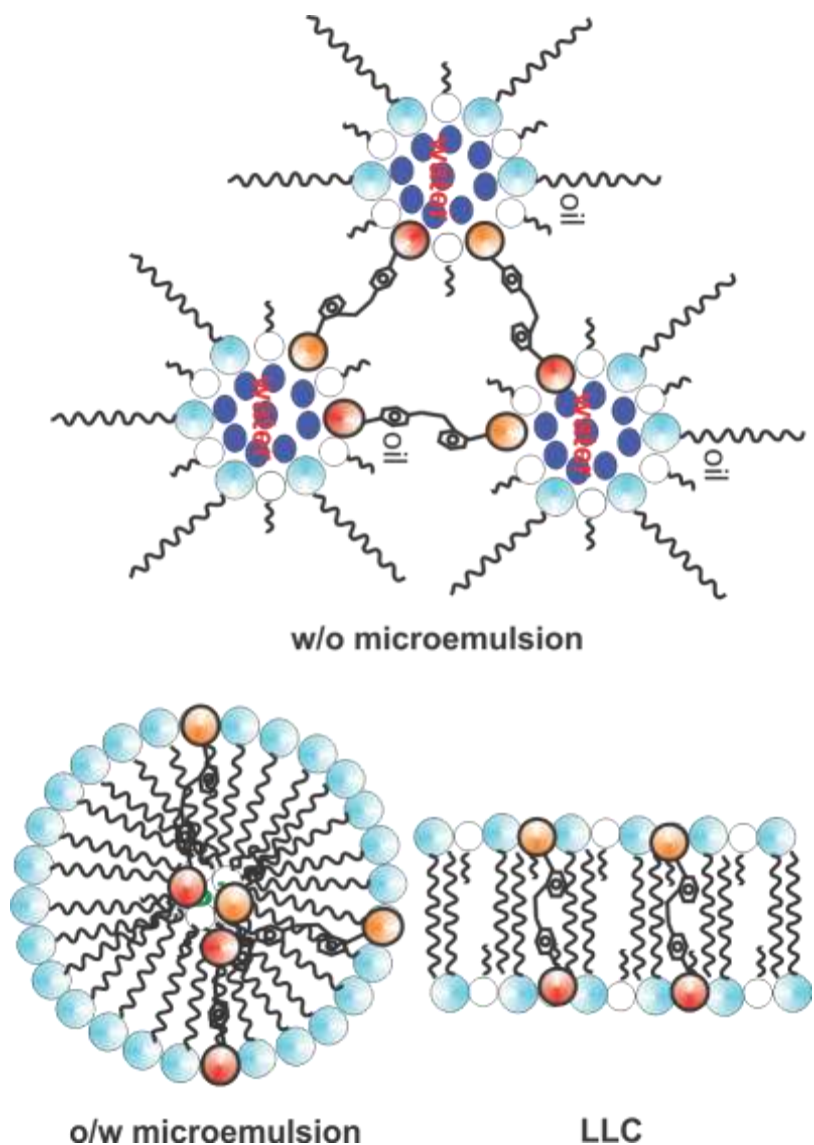

**Figure. 5. Chemical interaction between methyl red in the association structures**

Solubilisation of methyl red in LLC is more preferable due to the maximum interaction of polar head groups in LLC structure. Herein, the methyl red tail groups arranged in parallel position to the oil, this state associated with main LLC structure. Nevertheless, the solubility of methyl red would maximize in this model. In w/o microemulsion, the methyl red molecules arrange the **msd** structures associated together to form compact structures to help the dilution of methyl red. However, in o/w microemulsion, the existence of methyl red tends to break the globular structures. This process would not easy to be done, due to the associative forces of o/w microemulsion structure. Therefore, at o/w microemulsion a limited number of methyl red would dissociated in structure. The complete picture of the solubilization process could be seen in **Figure. 5**.

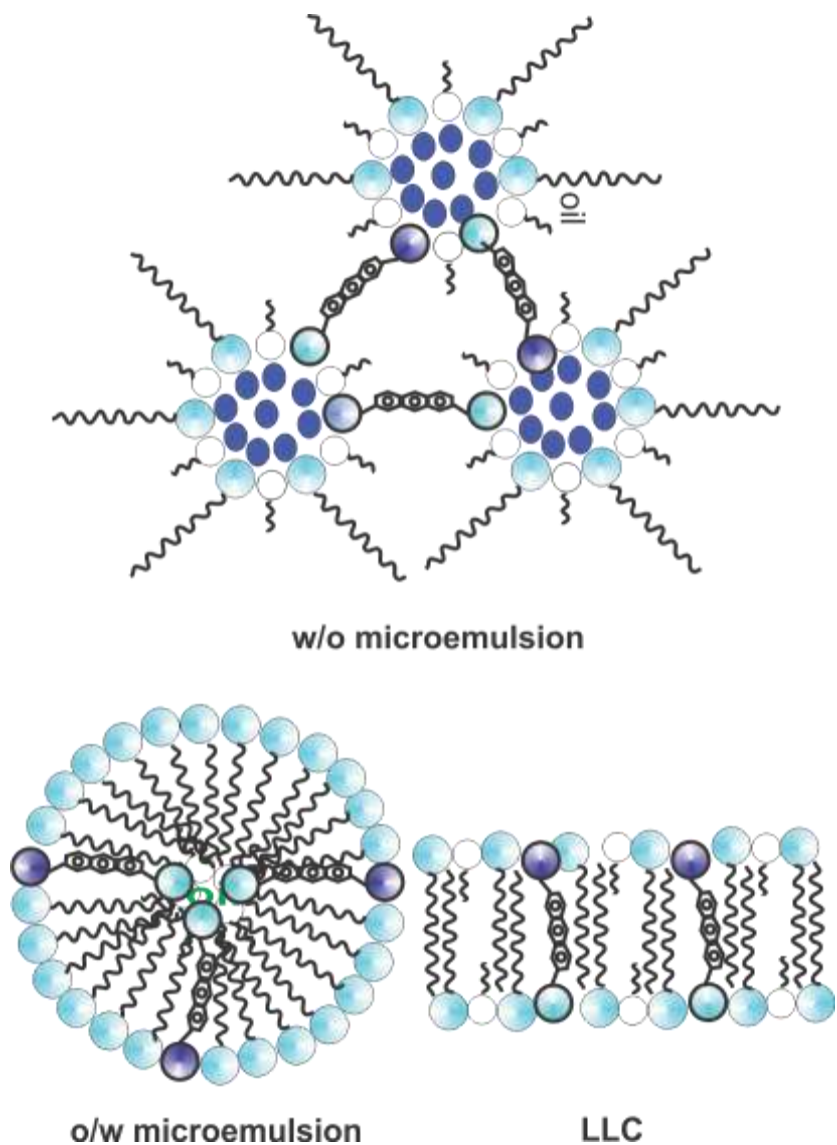

**Figure. 6. Chemical interaction between methylene blue in the association structures**

The solubilisation of methylene blue in association structure were unique, due to the methylene blue could almost be diluted in the number of proportions. As we could seen in **Figure.6**, the solubilisation of methylene blue in w/o microemulsion were preferable since the existence of methylene blue would associate **msd** to combined each other, so the chemical tension would be minimized. In addition, the chemical structure for methylene blue, a bit short in comparison to methyl red, so the chemical potentials of methylene blue could break the o/w structures and it could explained the high solubility of methylene blue in o/w structures. The limited solubility of methylene blue in LLC could be explained by: the short carbon chain would force the layer structures to distort in the given position where the methylene blue associates. Because of the tension, the solubility of methylene blue was restricted.

# CONCLUSIONS

In this research we have figured out the association structures for system of water at pH=4.5, and pH=9.5; SDS and penthanol. Area of w/o, o/w microemulsion, lamellar liquid crystal and hexagonal for water at pH=4.5 system were at coordinate positions around 12-60% water; 6-35% SDS; and 30-70% penthanol: 70-100% water; 0-25% SDS; and 0-14% penthanol: 23-55% water; 24-60% SDS; and 14-25% penthanol: 38-55% water; 40-57% SDS; and 1-7% penthanol contents, respectively. Whereas Area of w/o, o/w microemulsion, lamellar liquid crystal and hexagonal for water at pH=9.5 system were at coordinate positions around 12-60% water; 6-50% SDS; and 30-77% penthanol : 63-100% water; 0-26% SDS; and 0-14% penthanol : 25-60% water; 19-49% SDS; and 18-28% penthanol : 47-66% water; 28-48% SDS; and 0-8% penthanol contents, respectively.

The limited solubility for methyl red take place in o/w microemulsions since methyl red need an extra potential to break the **msd** structure in order to put methyl red molecules in. Whereas, the limited solubility for methylene blue take place in LLC due to the short carbon chain would force the layer structures to distort in the given position where the methylene blue associates.

# AKNOWLEDGEMENT

This research was supported by a grant Penelitian Program Desentralisasi Skema Hibah Bersaing Dana BOPTN Tahun Anggaran 2013 Kontrak No: 298.a.54/UN35.2/PG/2013 tanggal 15 Mei 2013.